\documentclass[twocolumn,aps,prl,superscriptaddress,amssymb]{revtex4}
\usepackage{graphicx,epsf,epsfig}
\usepackage{dcolumn}
\usepackage{bm}
\usepackage{color}
\usepackage{ulem}
\usepackage{subfigure}

\begin{document}

\title{Controlling transistor threshold voltages using molecular dipoles}
\author{Smitha Vasudevan}
\affiliation{Department of Electrical and Computer Engineering, University of Virginia, Charlottesville, VA 22904}
\author{Neeti Kapur}
\affiliation{Department of Chemical Engineering, University of Virginia, Charlottesville, VA 22904}
\author{Tao He}
\affiliation{Department of Chemistry, Rice University, Houston, TX 77005}
\affiliation{Smalley Institute for Nanoscale Science and Technology, Rice University, Houston, TX 77005}
\author{Matthew Neurock}
\affiliation{Department of Chemical Engineering, University of Virginia, Charlottesville, VA 22904}
\affiliation{Department of Chemistry, University of Virginia, Charlottesville, VA 22904}
\author{James M. Tour}
\affiliation{Department of Chemistry, Rice University, Houston, TX 77005}
\affiliation{Smalley Institute for Nanoscale Science and Technology, Rice University, Houston, TX 77005 }
\author{Avik W. Ghosh} 
\affiliation{Department of Electrical and Computer Engineering, University of Virginia, Charlottesville, VA 22904}
\widetext

\begin{abstract}
We develop a theoretical model for how organic molecules can control the electronic 
and transport properties of an underlying transistor channel to whose surface they 
are chemically bonded. The influence arises from a combination of long-ranged dipolar 
electrostatics due to the molecular head-groups, as well as short-ranged charge transfer 
and interfacial dipole driven by equilibrium band-alignment between the molecular backbone 
and the reconstructed semiconductor surface atoms. 
\end{abstract}
\maketitle

Inorganic semiconductors have traditionally dominated as the material players in the electronics industry. 
While their organic counterparts have been studied extensively as alternate channel materials \cite{reed},
the development of a stand-alone molecular electronics technology has been stymied by the inordinate 
difficulty of contacting small molecules reproducibly, their insufficient mobilities, large RC constants 
and poor gateability. Perhaps a more promising approach is to envisage hybrid organo-semiconductor devices, 
combining the established infrastructure of the semiconductor integrated industry with the `bottom-up' self-assembly 
and chemical tunability of molecular monolayers. A particularly interesting possibility is to use organic 
molecules to control the surface properties of deeply scaled, backgated silicon transistors, by tying up 
deleterious surface states and charge transfer `doping'. In addition, monitoring the transistor dynamics, 
such as the shift in its threshold voltage, can be used to detect a single molecule adsorption event \cite{lieber}. 
It is thus critical to properly understand the physical factors that determine how a molecule controls a 
transistor, and how the transistor, in turn, senses the molecule. 

In this paper, we develop a quantitative theory for the threshold voltage control of low-doped silicon channels 
by surface bonded organic monolayers with varying dipole moments. We focus in particular on a recent series of 
experiments \cite{taohe} that involved the grafting of molecular monolayers atop oxide free H-passivated silicon 
surfaces. The choice of the molecules followed an important logic. An identical set of molecules was used, with 
the exception of one substituent group. This allowed a systematic study of the effect of the molecules on the 
electrical properties of the device. While organic molecules attached to semiconductor surfaces have been 
studied extensively \cite{bocharov,vuillaume,cahen1}, albeit phenomenologically, our principal challenge is 
to develop a quantitative, `first principles' model that combines atomistic charge-transfer processes with macroscale 
electrostatics. We employ Density Functional Theory (DFT) to extract the molecular adsorption geometry, interfacial 
dipole and band-alignment at the atomistically reconstructed silicon surface. These quantities are then incorporated 
as inputs into a macroscopic Poisson solver to compute the band-bending in the transistor channel. The calculated 
threshold voltage shifts and band-alignments are in excellent agreement with experiments for multiple molecules 
(Table I) \cite{taohe}. 

Given the low doping and likely statistical fluctuations in data, as well as uncertainties in the experiment
(for instance, in the number of bonded molecules), there is undoubtedly some wiggle room for theory that may
make this correspondence fortuitous. However, the main trend in the experiment seems quite robust, namely, a
clear monotonic dependence of the threshold voltage on the dipole moment of the headgroup. This is borne out
quite well by our model. In addition, we demonstrate the accuracy of our model by comparing our computed 
intermediate ingredients with complementary experimental data obtained on the same set of samples \cite{taohe_surface}. 

\begin{figure}[h]
\hskip 0.2cm\centerline{\epsfxsize=3.3in\epsfbox{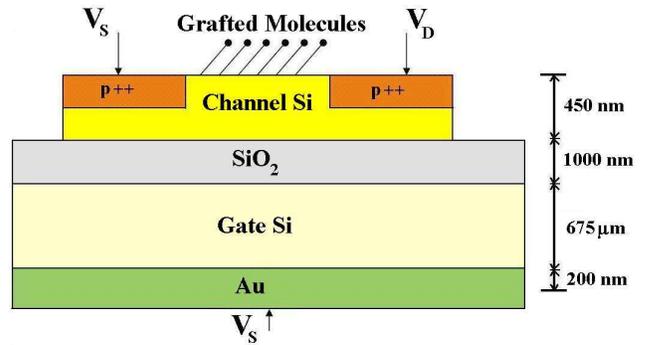}}
\caption{Schematic side-view representation (not to scale) of the pseudo-MOSFET under investigation 
\cite{taohe}. The molecules were grafted between the source and drain electrodes. $V_S$, $V_D$, and $V_G$ 
refer to the bias applied on the source, drain, and gate, respectively.}
\label{schematic}
\end{figure}

\begin{figure}[h]
\hskip 0.1cm\centerline{\epsfxsize=3.3in\epsfbox{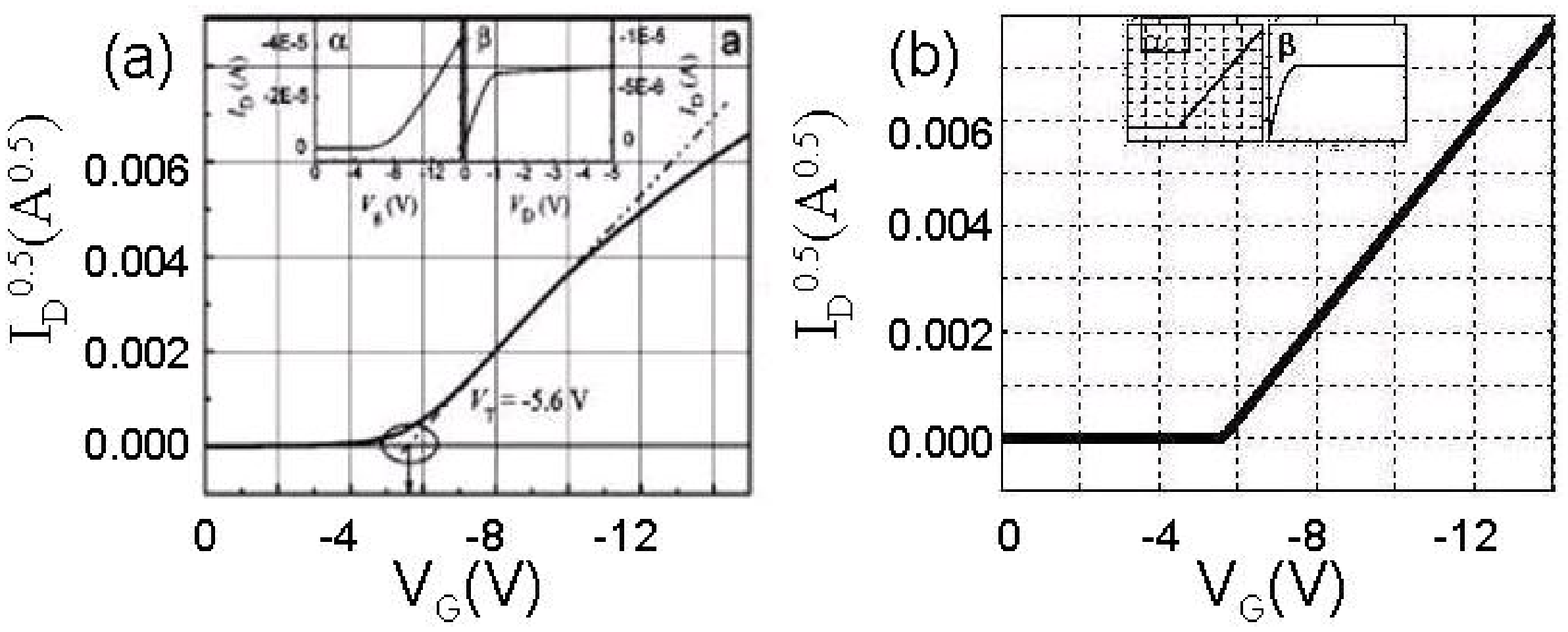}}
\caption{(a)Experimentally obtained I-V characteristics \cite{taohe} of the fabricated MOSFET devices. Inset $(\alpha )$ displays the typical transfer characteristics of the devices under test. Inset $(\beta )$ shows the typical output characteristics of the devices under test.(b)Theoretically calculated I-V characteristics of the MOSFET device.}
\label{bareiv}
\end{figure}

{\it{Molecule free device.}}
The devices we are studying were fabricated using a silicon-on insulator (SOI) wafer \cite{taohe}. A back 
gated structure was used to allow easy grafting of a molecular monolayer on the silicon device top layer (Fig.\ref{schematic}). 
The molecules were grafted on H-passivated silicon surfaces using diazonium based chemistry (molecules lose the diazonium group to form radicals before attaching to the surface\cite{diazo}). In the fabricated MOSFET, the bulk p-Si substrate (handle) was biased at $V_G$ using the 
gold back contact and acts as a gate terminal. This in turn induces a conduction channel at the upper 
interface of the buried oxide, used as a gate dielectric layer. The conducting channel corresponds to the 
formation of an accumulation layer (p-channel). In practice, the drain current voltage characteristic presents 
a linear behavior at low drain voltages $V_D$, and tends to saturate as the drain voltage approaches the voltage difference between the gate bias and the threshold voltage. In the diffusive transport regime, the current-voltage characteristics are satisfactorily described by 
the square law theory \cite{sze}(Fig.\ref{bareiv})
\begin{equation}
I = \frac{qC_{G}\mu W}{2L}\Biggl[V_G-V_T-\frac{V_D}{2}\Biggr]V_D
\end{equation}
where $W$ and $L$ are the channel width and length, $C_{G}$ is the
gate capacitance and $\mu$ is the channel mobility. In this case, an accumulation channel is activated when 
both the gate and the drain are negatively biased. Because the transistor body is the nearly intrinsic p-Si 
layer, the channel is assumed to be completely accumulated. 

An accumulation mode transistor is not expected to have a prominent turn-on as one expects for an
inversion channel transistor. The onset voltage is associated with pulling the contact Fermi energies
below the valence band-edge near the oxide interface. The large threshold voltage values of the bare MOSFET 
($\sim-5.6 V$) could be indicative of traps at the oxide-channel interface. This is further confirmed by the presence of
prominent hysteretic signatures that suggest charge trapping processes.
We used two approaches to estimate the areal density of interfacial traps in the control devices. The 
first assumes that the threshold voltage of the control devices is due to trapped charge on one side of a 
capacitor, with a capacitance/area equal to $C_{ox}$. Using this method, the density of interfacial traps 
was calculated to be $N_{i}=Q_F/q=|\Delta V_T|\times C_G/{q}=1.2\times 10^{11}/cm^2$. The second approach
estimates the same quantity using the observed subthreshold swing S \cite{rolland}. The traps would degrade 
the subthreshold current by taking away some of the mobile charges that would otherwise have contributed to 
current flow. The interfacial trap density is calculated to be $N_{i}=SC_G\log_{10}e/(kT/q)=1.0\times10^{11}/cm^2$. The consistency 
of the two estimates strengthens our arguments as to the origin of the threshold voltage and subthreshold slopes 
observed. 

\begin{figure}[h]
\hskip 0.2cm\centerline{\epsfxsize=3.3in\epsfbox{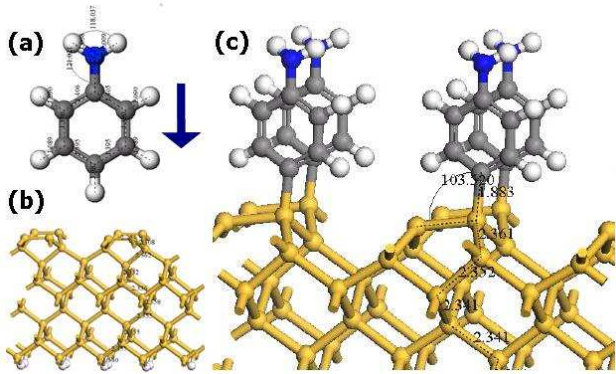}}
\caption{Optimized geometries for the (a)gas-phase molecule aniline(arrow points in the direction of dipole 
moment, from positive to negative charge).(b)Optimized structure for passivated Si(100) slab with symmetrically 
dimerized surface.(c) Optimized structure for aniline radical adsorbed on passivated Si(100)-(2x1) slab atop 
site T1( top atom of the dimer).} 
\label{optim}
\end{figure}

{\it{Molecular attachment.}}
Quantum-chemical calculations were carried out using DFT as implemented in the
Vienna Ab-initio Simulation Package (VASP) \cite{vasp1,vasp2}. Plane wave basis sets were used to represent 
the atoms and the PW91 form of the GGA functional was used to carry out gradient corrected calculations.
The optimized structures for the isolated gas-phase molecules under consideration, dimethylaminobenzene, aniline and 
nitrobenzene and the corresponding radical forms are shown in Fig.\ref{optim}. These structures were obtained using spin-polarized calculations, and are in agreement 
with published experiments and theoretical computations \cite{tzeng,palafox}.

The Si(100) slab was modeled using 9 atomic layers and hydrogen atoms were attached on the lowest layer of 
the slab to passivate the dangling bonds on the lower slab surface. Geometry optimizations were carried out 
while constraining the lowest 4 layers of the Si(100) slab at their bulk positions including the hydrogen atoms 
attached on the lowest atomic layer. We started with a dimerized Si(100) structure as our initial guess
(with broken symmetry), and recovered the expected 2x1 reconstructed structure.
The dipole moments and vibrational modes were calculated within VASP using single point calculations.

Once our isolated geometries were thoroughly benchmarked, we moved on to the adsorption geometry. The radical forms of 
the molecules we studied, specifically, dimethylaminobenzene, aniline and nitrobenzene, were adsorbed on the 
asymmetrically dimerized Si(100) slab via a C-Si covalent bond. We considered two possible adsorption sites on the 
Si(100) slab, named T1 and T2. In all the cases 
considered for atop adsorption of radicals on the surface, we find that the radicals stand normal to the surface.
A positive dipole, with the positive pole directed toward the semiconductor surface, is induced by nitrobenzene, 
while a negative dipole is induced by aniline and dimethylaminobenzene. Table I shows the computed dipole moments.

\begin{figure}[h]
\hskip 0.2cm\centerline{\epsfxsize=3.3in\epsfbox{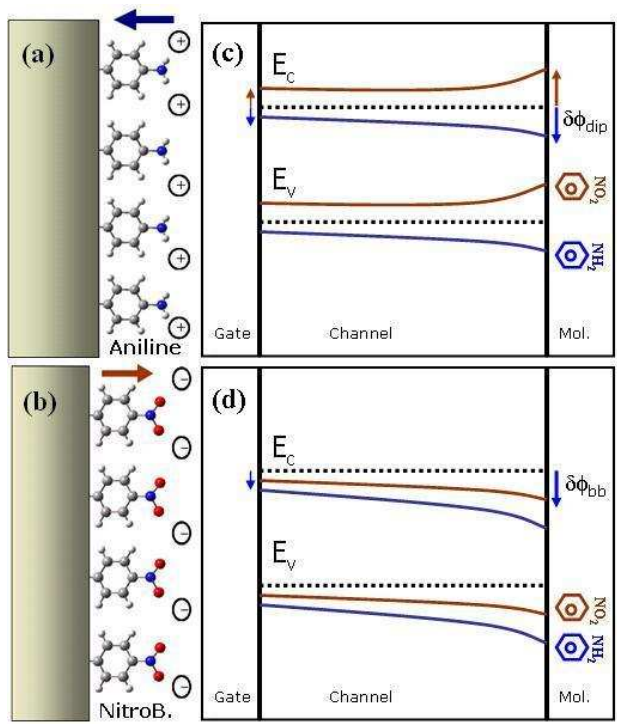}}
\caption{Adsorbed layer of (a)Aniline (Blue line) and (b)Nitrobenzene (Brown line) molecules, creating dipole moments on the surface. (c) Energy Band Diagram for the system being studied.(d) Pictorial representation of change in band bending due to charge transfer between molecular monolayer and substrate.}
\label{dip_bb}
\end{figure}

{\it{Threshold Voltage Shift.}}
The presence of the molecular layer modifies the silicon surface work function, which in turn affects the device 
properties. The work function (WF) is the minimum energy required for an electron to escape into vacuum from the 
Fermi level ($E_F$) of the material, and is determined by (i) the electron affinity (EA), the energy required to 
excite an electron  from the bottom of the conduction band (CB) at the surface to the local vacuum level; (ii) 
the band bending (BB), the electrical potential difference between the surface and the electrically neutral 
semiconductor bulk; and (iii) the energy difference between the bulk CB and the Fermi level. $\Delta$WF can be due 
to $\Delta$EA, $\Delta$BB, or both. We computed the total shift directly from the shift in the estimated highest
occupied molecular orbital (HOMO) level of the molecule before and after adsorption. This was accomplished by
comparing the projected density of states on the silicon atoms, which is maximized for the HOMO levels of 
benzene-based molecules \cite{damle}.
\begin{figure}[h]
\hskip 0.2cm\centerline{\epsfxsize=3.5in\epsfbox{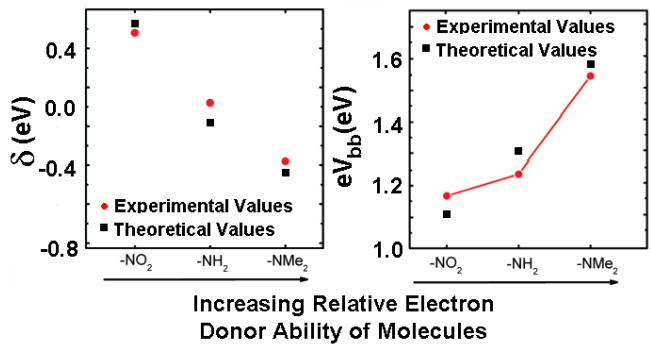}}
\caption{Comparison of calculated values and experimental UPS/IPES/XPS data. Experimental data are adapted from \cite{taohe_surface}} 
\label{ups}
\end{figure}
The computed shifts agree very well with experimental data, as seen from the last two columns of Table I, as well
as comparisons with ultraviolet photoelectron spectroscopy (UPS)and inverse photoemission spectroscopy (IPES) data combined
with X-ray photoelectron spectroscopy (XPS) data (Fig.\ref{ups}) The shift
shows a monotonic dependence on the dipole moment of the headgroup, which is quite encouraging. Surprisingly, 
however, there is an additional shift relative to the H-passivated control, which is unexpected (thus, the threshold
voltage shifts of the nitro and amino components do not straddle that of the control, but lie on the same side of it). 
We will now attempt to deconstruct the computed shifts to understand their physical origins. 
{\it{Dipolar Contribution.}}
The electron affinity of a given surface is directly affected by a surface dipole, which creates an electrical 
potential drop across the grafted molecular film depending on the dipole moment. Simply put, a positive headgroup
such as $NH_2$ acts as a gate that pulls negative charges to the surface, lowering its local electronic levels,
while $NO_2$ pushes them away (Fig.\ref{dip_bb}). The resulting band-bending at the surface propagates towards the opposite
end where the oxide sits, penetrating up to a Debye length $L_D = \sqrt{\epsilon_r\epsilon_0k_BT/N_Aq^2}$, where 
$\epsilon_r$ is the dielectric constant of the channel, $\epsilon_0$ is the permittivity of free space, $T$ is the 
temperature and $N_A$ is the acceptor dopant density.  

\begin{table*}[ht]
\caption{Data of computed dipoles of the molecule-semiconductor system, shifts due to dipole ($\phi_{dip}$) and charge 
transfer ($\phi_{bb}$), computed and experimental changes \cite{taohe} in the threshold voltage of the devices. The 
control consists of an H-passivated Si(100) surface.}
\centering
\begin{tabular}{c c c c c c}
\hline \hline
{Systems} & $\mu_{dip}$ & $\delta\phi_{dip}$ & $\delta\phi_{bb}$ & $\Delta V_T(V)$ &  $\Delta V_T(V)$\\ 
 &  (dB) & (V) & (V) &Theory & Expt.\\ \hline

\hline
Control & 0.03 & 0.0058 & 0 & 0.0058 & 0 \\
Nitro & 2.17 & 0.22 & -0.725 & -0.525 & -0.55 \\
Aniline & -1.44 & -0.14 & -0.96 & -1.1 & -1.25 \\ 
Dimethylaminobenzene & -1.65 & -0.28 & -1.12 & -1.4 & -1.55 \\ [1ex]
\hline
\end{tabular}
\label{table}
\end{table*}
The purely dipolar headgroup gives a shift in potential given by 
\begin{equation}
\delta\phi_{dip} = 
(N_{mol}\mu_{dip}\cos{\theta}/\epsilon_m)e^{-t_{si}/L_D}
\end{equation}
where $N_{mol}$ gives the areal surface density of the attached molecules, $\mu_{dip}$ is the static dipole 
moment of each molecular head group oriented at an average angle $\theta$ relative to the normal to the surface, $\epsilon_m$ is 
the molecular dielectric constant, and $t_{si}$ is the thickness of the channel silicon. This contribution is opposite for 
opposite dipolar signs. Significantly, dipolar gating at the top surface nontrivially influences the threshold 
voltage at the backgate owing to the large Debye length associated with low doping in the channel. For a doping 
level of $10^{13}cm^{-3}$ the Debye length $L_D \sim 1.25 \mu m$, so that the exponential transfer factor is
around 0.7 (implying that seventy per cent of band-bendings at the surface are transmitted to the bottom end).

{\it{Charge-Transfer Barrier.}}
In addition to the dipolar contribution, the molecular monolayers also transfer spectral weight to the semiconductor
surface through bonding. Because the mobile charges are primarily near the bottom of the channel, this does not add
significant additional resistance, but it does influence the local band-bending through the charge-transfer dipole
at the molecule-semiconductor interface. Based on
our computed shifts and the dipolar contributions separately extracted, we can compute this component for each 
molecule. To zeroth order, this represents the charge transfer due to the workfunction difference between the 
headgroup-free molecular backbone (the bare benzene ring) and the silicon surface. This means that the associated
band-bending is in the same direction for all the headgroups (Fig.\ref{dip_bb}b), and relatively weakly varying with dipole
moment (column 4 of Table I). There is, however, some variation, indicating that the separation into a purely 
dipolar part and a head-group independent backbone part is not strictly feasible, owing to the short length of the
benzene rings and the resulting hybridization between the ring states and the headgroups. The electron donating 
ring electrons push charge towards the p-Si substrates and become partly positively charged, leading to an 
increase in the positive surface charge density and a lowering of the local silicon surface levels (Fig.\ref{dip_bb}b). The
donating capacity, however, is enhanced (diminished) by the electronegativity (positivity) of the headgroups. 
Methylamine has the highest electron-donation capability, and hence provides the highest change in charge-transfer
induced band bending $\delta \phi_{bb}$. While we do not independently compute $\delta \phi_{bb}$, one can roughly
estimate this by calculating the charge neutrality level $E_{CNL}$  of the headgroup molecular backbone (CNL is defined 
as the energy to which electrons need to fill the molecule to keep it electrically neutral -- frequently this is
replaced by the LUMO level \cite{cahen1}). The work-function 
shift of the molecule is then given by 
\begin{equation}
\delta\phi_{bb}\approx \frac{(E_F-E_{CNL})e^{-t_{ch}/L_D}}{1+1/U_0D_0}
\end{equation}
where $U_0$  is the single-electron charging energy of the molecule and $D_0$  is its density of states near the 
Fermi energy. The barrier height $E_F - E_{CNL}$ is of the order of 1 V for aromatic molecules on silicon \cite{cahen1},
giving us the correct observed order of magnitude. The term in the denominator includes the Coulomb cost for charging 
up the molecule. 
In this contribution, we have studied the effects, both qualitatively and quantitatively, of grafting 
molecular monolayers on the top surface of back-gated MOSFETS. Such a scheme of using molecules on surfaces 
to tailor device properties is becoming increasingly attractive as devices get smaller, and surface-to-volume 
ratios increase vastly. The study of surface modifications could also have practical implications for sensors, 
specifically chemically sensitive field-effect transistors (CHEMFETs)\cite{nikolaides}. A radically different 
principle of operation that we are working on \cite{ieee} involves a stronger bonding between the molecule and the 
channel, involving actual wavefunction overlap, leading to quantum scattering by the molecular traps that creates 
characteristic fingerprints when scanned with a back gate.

We would like to acknowledge discussions with Tao He, Jim Tour, Lloyd Harriott, John Bean, Neil Di Spigna, Paul Franzon, Pradeep Nair and Ashraf Alam. This work was supported
by the NSF-NIRT and NSF-CAREER awards under grants GA10646-128609 and GA10696-129495 respectively.

\end{document}